\documentclass[1p,12pt]{elsarticle}

\usepackage[version=4]{mhchem}
\usepackage{graphicx}
\usepackage{amsmath,amssymb}
\usepackage{array,multirow}
\usepackage{longtable}
\usepackage{xcolor}
\usepackage[normalem]{ulem}
\usepackage{hyperref}

\journal{Materials Science \& Engineering B}

\begin{document}

\begin{frontmatter}

\title{Metallic Organometallic and Semiconducting Covalent Phases of Free-Standing \texorpdfstring{$\gamma$}{gamma}-Graphdiyne Molecular Wires}

\author[ppgee-unb,nanoeng-unb]{A. G. L. Rodrigues}
\author[unicamp]{G. S. L. Fabris}
\author[unicamp]{B. Ipaves}
\author[ppgee-unb,ene-unb]{\\F. L. L. de Mendon\c{c}a}
\author[unicamp]{D. S. Galv\~ao}
\author[ppgee-unb,nanoeng-unb,ene-unb]{M. L. Pereira Junior\corref{cor1}}
\ead{marcelo.lopes@unb.br}
\cortext[cor1]{Corresponding author.}

\affiliation[ppgee-unb]{organization={Postgraduate Program in Electrical Engineering, College of Technology, University of Bras\'ilia},%
   postcode={70910-900}, city={Bras\'ilia}, state={Federal District}, country={Brazil}}
\affiliation[nanoeng-unb]{organization={NanoEngineering Laboratory, College of Technology, University of Bras\'ilia},%
   postcode={70910-900}, city={Bras\'ilia}, state={Federal District}, country={Brazil}}
\affiliation[unicamp]{organization={Applied Physics Department, ``Gleb Wataghin'' Institute of Physics, State University of Campinas},%
   postcode={13083-859}, city={Campinas}, state={S\~ao Paulo}, country={Brazil}}
\affiliation[ene-unb]{organization={Department of Electrical Engineering, College of Technology, University of Bras\'ilia},%
   postcode={70910-900}, city={Bras\'ilia}, state={Federal District}, country={Brazil}}

\begin{abstract}
One-dimensional carbon allotropes that combine $sp$ and $sp^{2}$ hybridizations are an actively pursued platform for atomically precise wires whose electronic, vibrational, and mechanical responses can be tuned by chemical substitution along the backbone. Recent on-surface synthesis has realized one such platform, namely the $\gamma$-graphdiyne molecular wire on Au(100), in two interconvertible phases of identical aromatic backbone but distinct linkage chemistry, an organometallic intermediate (OMW) with $\mathrm{C{-}Au{-}C}$ bridges and a fully covalent product (COW) with diacetylenic linkages, with the on-surface structural periodicity and the $\mathrm{C}{\equiv}\mathrm{C}$ Raman signature characterized experimentally. The intrinsic properties of the free-standing wires, however, remain inaccessible to the surface-supported measurement, and we address this gap with a hybrid-functional first-principles characterization of both phases. The COW phase is a one-dimensional semiconductor with a direct band gap of $E_{g} = 2.02$~eV, a polyynic bond-length alternation of $0.117$~\AA, and a chain periodicity that reproduces the experimental value within $-0.5\%$. The OMW phase is partially cumulenized and is found to be a one-dimensional metal, with the spin channels of the broken-symmetry hybrid solution crossing the Fermi level along the chain Brillouin zone. Both polymers are dynamically stable as isolated objects, with one-dimensional Young's moduli of $995$~nN and $861$~nN respectively, and the $\mathrm{Au{-}C}$ bridge is identified as a single-atom soft link that absorbs $9.2\%$ of the axial deformation at $+6\%$ macroscopic strain. Species-resolved vibrational analysis isolates an effective-conjugation-coordinate band at $1408$~cm$^{-1}$ as a fingerprint of COW and a manifold of Au-projected modes below $400$~cm$^{-1}$ as the analogous fingerprint of OMW, providing diagnostic signatures for on-surface phase discrimination during the OMW-to-COW conversion.
\end{abstract}


\end{frontmatter}

\section{Introduction}

Carbon-based materials occupy a central position in materials science and in molecular nanotechnology, in part because the three native hybridization states of carbon, namely \textit{sp}, \textit{sp}$^{2}$, and \textit{sp}$^{3}$~\cite{Hirsch2010}, which enable a structural versatility that has been translated into atomically precise architectures with electronic, vibrational, and mechanical responses tuned to specific functions. The discovery of fullerenes~\cite{Kroto1985}, the observation of carbon nanotubes~\cite{Iijima1991}, and the isolation of monolayer graphene~\cite{Novoselov2004} demonstrated the experimental accessibility of low-dimensional carbon allotropes and triggered an intense investigation of their physical properties and of their integration into single-molecule and few-molecule devices. Beyond these emblematic systems, theoretical predictions advanced by Baughman \textit{et al.} in 1987~\cite{Baughman1987} introduced graphynes and graphdiynes as a class of two-dimensional networks combining \textit{sp} and \textit{sp}$^{2}$ carbon atoms in periodic arrangements, paving the way for a tunable family of allotropes whose acetylenic linkages confer non-zero electronic band gaps and intrinsic structural porosity of immediate interest to charge-transport, gas-sensing, and energy-storage applications~\cite{Huang2018, Xue2018, LiYL2014}. The synthesis of graphdiyne films by Li \textit{et al.} in 2010~\cite{LiYL2010} confirmed the experimental viability of these structures and motivated sustained efforts toward additional members of the graphyne family, including the scalable solution-phase synthesis of multilayer $\gamma$-graphyne crystals~\cite{Desyatkin2022} and a planar-sheet $sp^{2}$ carbon phase derived from $\gamma$-graphyne at low temperature~\cite{Aliev2025}. We note, however, that the characterization underlying some of the proposed routes to $\gamma$-graphyne has been examined by independent replication and structural-fingerprint studies that reach more reserved conclusions about the identity of the products obtained~\cite{Kone2025,Martin2024}. Linear analogs containing only \textit{sp} carbon, namely polyynes and the elusive carbyne~\cite{Casari2016, Tykwinski2010}, complete this taxonomy and define a continuum of \textit{sp}/\textit{sp}$^{2}$ ratios with electronic, vibrational, and mechanical responses that depend systematically on the bonding network.

Among the graphdiyne polymorphs, the so-called $\gamma$-graphdiyne ($\gamma$-GDY), composed of benzene rings interconnected by diacetylenic ($\mathrm{-C{\equiv}C{-}C{\equiv}C-}$) linkages in a hexagonal lattice, has attracted considerable attention due to its predicted direct electronic band gap, high carrier mobility, and uniformly distributed nanopores~\cite{Huang2018, Xue2018}. First-principles studies have characterized the electronic structure of $\gamma$-GDY nanoribbons as a function of width and edge termination~\cite{Pan2011, Bai2011}, the modulation of the band gap by transverse electric fields~\cite{Kang2012}, and the consequences of isoelectronic boron and nitrogen substitutions on conjugation and charge distribution~\cite{Bu2012, Zhou2011}. The one-dimensional limit, in which a single $\gamma$-GDY chain forms an extended polymeric wire, has also been considered theoretically through the crystal orbital framework, indicating tunable electronic responses upon stretching and substitutional modification~\cite{Zhu2016}. Theoretical work on the underlying \textit{sp}-carbon backbone has further characterized the dependence of the bond-length alternation (BLA) and of the effective conjugation coordinate (ECC) mode frequency on chain length, hybridization, and termination chemistry~\cite{Yang2008, Cinquanta2011, Ravagnan2009}. Together, these studies position $\gamma$-GDY as a candidate building block for atomically precise molecular electronics, in which the polymeric wire serves as a one-dimensional channel whose response to local fields, strain, and termination chemistry can be engineered with chemical precision, while the truly one-dimensional limit, in which the bare polymeric wire is isolated as a free-standing object, remains comparatively underexplored from a first-principles perspective.

The bottom-up assembly of carbon nanostructures via on-surface synthesis under ultra-high vacuum has emerged as a controllable route to atomically precise low-dimensional architectures~\cite{Cai2010, Bjork2013, Bjork2016}. The Ullmann coupling of halogenated aromatic precursors on noble metal surfaces, in particular, allows the formation of organometallic intermediates that, upon thermal annealing, evolve into fully covalent organic networks~\cite{Bjork2013, Klappenberger2018, DeBoni2020a}. This protocol has been extended to graphyne-like and graphdiyne-like architectures, with notable demonstrations of single-layer molecular networks on Au(111)~\cite{Rabia2020, Achilli2021} and one-dimensional molecular wires obtained from di-bromo-aryl alkynyl precursors~\cite{Sedona2020, DeBoni2024, Klappenberger2018}. Very recently, Cartoceti \textit{et al.}~\cite{Cartoceti2026} reported the surface-orientation-dependent synthesis of $\gamma$-GDY molecular wires from 1,4-bis(bromoethynyl)benzene precursors on Au(100) and Au(111), demonstrating that both the organometallic intermediate, in which $\mathrm{C{-}Au{-}C}$ bridges interrupt $\pi$-conjugation, and the fully covalent diacetylenic wire are stable on the substrate at controlled temperatures. The two phases, therefore, constitute the same precursor processed into two distinct one-dimensional polymers with potentially different electronic and mechanical responses, providing an experimental platform for probing how the insertion of a single metal atom into an otherwise $\pi$-conjugated wire reshapes its physical properties.

Despite these experimental advances, three central questions about the intrinsic behavior of the two phases remain unanswered. First, the intrinsic properties of the free-standing one-dimensional polymer are inaccessible to on-surface measurements, which mix the chain response with substrate-induced renormalization and limit vibrational diagnostics to zone-center Raman spectra of substrate-decoupled fragments. Second, the electronic ground state of the organometallic phase has not been resolved at the hybrid-functional level with explicit treatment of the spin-symmetry-breaking solutions, for which generalized gradient approximation (GGA) functionals are known to have problems due to their over-delocalization error~\cite{Cohen2008,MoriSanchez2008}. Third, the response of each phase to axial deformation, including the dependence of the COW gap on strain through the bond-length alternation and the mechanical role of the $\mathrm{Au{-}C}$ bridge in OMW, has not been quantified for the free-standing wires.

In this work, we present a first-principles characterization of the free-standing $\gamma$-graphdiyne molecular wire in both its organometallic and covalent phases, addressing the three open questions mentioned above. We combine PBE+MBD geometry optimizations, HSE06 hybrid-functional electronic structure with explicit spin-symmetry breaking, finite-displacement Phonopy lattice dynamics, species-resolved mode decomposition, and a three-layer strain protocol that spans $\varepsilon \in [-6, +6]\%$ at the atomic, electronic, and vibrational levels. The free-standing dispersion is consistent with the intrinsic dynamical stability of both phases, the species-resolved fingerprints identify diagnostics for on-surface phase discrimination beyond the $\mathrm{C}{\equiv}\mathrm{C}$ shift alone, and the dense-path HSE06 sampling resolves the electronic ground state of OMW as a metallic configuration in which the spin-$\alpha$ and spin-$\beta$ channels of the collinear ferromagnetic solution cross the Fermi level at distinct wave vectors. Beyond the specific system, the analysis exposes a methodological pitfall transferable to the broader class of $\pi$-conjugated wires functionalized with single-atom metal bridges, in which the small direct gap reported by $k$-mesh-based estimators in single-point hybrid calculations does not survive a dense $k$-path sampling and reflects the discrete spacing between mesh points on opposite sides of the spin-resolved Fermi crossings rather than a physical band gap.

\section{Methodology}

All first-principles calculations were carried out with the all-electron numerical-atomic-orbital code FHI-aims~\cite{Blum2009} at the tight numerical tier, using the PBE functional~\cite{Perdew1996} with the many-body dispersion (MBD) correction in the MBD@rsSCS variant~\cite{Tkatchenko2012,Ambrosetti2014}. Gold was treated within the scalar atomic zeroth-order regular approximation (ZORA). Spin-orbit coupling was not included, since the states near the Fermi level are dominated by carbon $sp$ and $sp^2$ orbitals. Hybrid-functional refinements used HSE06 with the parameters of Krukau \textit{et al.}~\cite{Krukau2006} ($\omega = 0.11~\mathrm{bohr^{-1}}$, $25\%$ Hartree--Fock admixture) as single points on the relaxed PBE+MBD geometries, preserving the self-consistent field (SCF) tolerances and the $k$-mesh of the PBE+MBD step.

The two polymeric phases were modeled as free-standing 1D wires with $30$~\AA{} of vacuum padding along the non-periodic directions, with the chain axis aligned with $\mathbf{c}$. The covalent organic wire (COW, $\mathrm{-[C_6H_4{-}(C{\equiv}C)_2]_n-}$) has $14$ atoms per primitive unit cell, whereas the organometallic wire (OMW, $\mathrm{-[C_6H_4{-}C{\equiv}C{-}Au{-}C{\equiv}C]_n-}$) has $15$ atoms per unit cell and one Au atom (5d$^{10}$6s$^1$) per formula unit, so that the elementary OMW unit has odd valence-electron parity and is incompatible with a closed-shell description. A four-candidate screening (closed-shell 2-unit, metallic 1-unit with $0.05$~eV Gaussian smearing, ferromagnetic 1-unit with $\mu_\mathrm{Au}^\mathrm{init} = +1$, antiferromagnetic 2-unit with $\pm 1$ alternating moments) returned converged energies within $1$~meV of each other and indistinguishable relaxed geometries, and the 1-unit ferromagnetic configuration was retained as the structural setup because it carries the spin-resolved Fermi crossings of the OMW phase under HSE06 (Section~S5). Finite-oligomer references used hydrogen termination for COW ($n = 1$, 1,4-diethynylbenzene) and bromine termination for OMW ($n = 2,3,4,6$, matching the precursor used in the on-surface synthesis of Cartoceti \textit{et al.}). The odd-electron $n = 2, 4, 6$ systems were treated as doublets and $n = 3$ as closed-shell.

Atomic positions were optimized with a trust-region quasi-Newton scheme to a maximum residual force of $5 \times 10^{-3}$~eV/\AA, with the chain-axis lattice vector allowed to relax and the two non-periodic vectors fixed. Periodic systems used a $1 \times 1 \times 24$ Monkhorst--Pack mesh, whereas finite oligomers were treated as isolated molecules. SCF tolerances were $10^{-6}$~eV on the total energy, $10^{-5}$ on the density, and $10^{-4}$~eV/\AA{} on the residual force, with $0.01$~eV Gaussian smearing for gold-containing systems. HSE06 band structures along $\Gamma \to \mathrm{X}$ were sampled with $101$ wave vectors to resolve the spin-resolved Fermi crossings of the OMW phase. For the strained OMW configurations at $\varepsilon = +3\%$ and $+6\%$, as well as for the OMW oligomer at $n = 6$, the hybrid evaluation was performed at the light tier as a controlled fallback, with a cross-check at $\varepsilon = 0$ confirming agreement on the Fermi-crossing wave vectors to the resolution of the $101$-point path (Table~S2).

Phonon dispersions were obtained by finite displacements in Phonopy~\cite{Togo2015} on $1 \times 1 \times 6$ supercells of both polymers, with the force threshold increased by an order of magnitude relative to the relaxation criterion. The $1 \times 1 \times 4$ convergence test on COW agrees within $0.4$~cm$^{-1}$ on every mode along $\Gamma \to \mathrm{X}$ (Table~S1). Finite oligomers were treated with $\Gamma$-only frequencies on isolated molecules. The harmonic spectrum confirms the dynamical structural stability and supplies zero-point energies, and the temperature dependence of $C_V$ and of the vibrational free energy are reported in the Supporting Information. Axial strain was mapped on the thirteen-point grid $\varepsilon \in \{-6, -5, \ldots, +5, +6\}\%$ in three layers. The first layer is the PBE+MBD relaxation under $c(\varepsilon) = (1 + \varepsilon)\,c_0$ at all thirteen points. The second layer is the HSE06 single-point evaluation on the relaxed geometries at the subset $\varepsilon \in \{-6, -3, 0, +3, +6\}\%$. The third layer is the $\Gamma$-only finite-displacement phonon calculation on every strained primitive cell.

The species-resolved decomposition of the vibrational modes was performed in post-processing on the Phonopy eigenvectors as a mass-weighted participation ratio:
\begin{equation}
p_\alpha(\nu, q) = \frac{\sum_{i \in \alpha} m_i |\mathbf{u}_i^{(\nu, q)}|^2}{\sum_i m_i |\mathbf{u}_i^{(\nu, q)}|^2},
\label{eq:participation}
\end{equation}
\noindent with species index $\alpha \in \{\mathrm{Au}, \mathrm{C}_{sp}, \mathrm{C}_{sp^2}, \mathrm{H}, \mathrm{Br}\}$, $m_i$ and $\mathbf{u}_i^{(\nu, q)}$ the mass and eigendisplacement of atom $i$, and $\sum_\alpha p_\alpha = 1$ in all modes. Carbon atoms were assigned to $sp^2$ when they had three or more carbon neighbors within $1.65$~\AA{} or one hydrogen neighbor within $1.20$~\AA, and to $sp$ otherwise. This topological criterion is stable across the full strain range and avoids the spurious bond-type reassignments produced by length-based schemes at intermediate strains.

\section{Results and Discussion}

\begin{figure}[htb!]
\centering
\includegraphics[width=\linewidth]{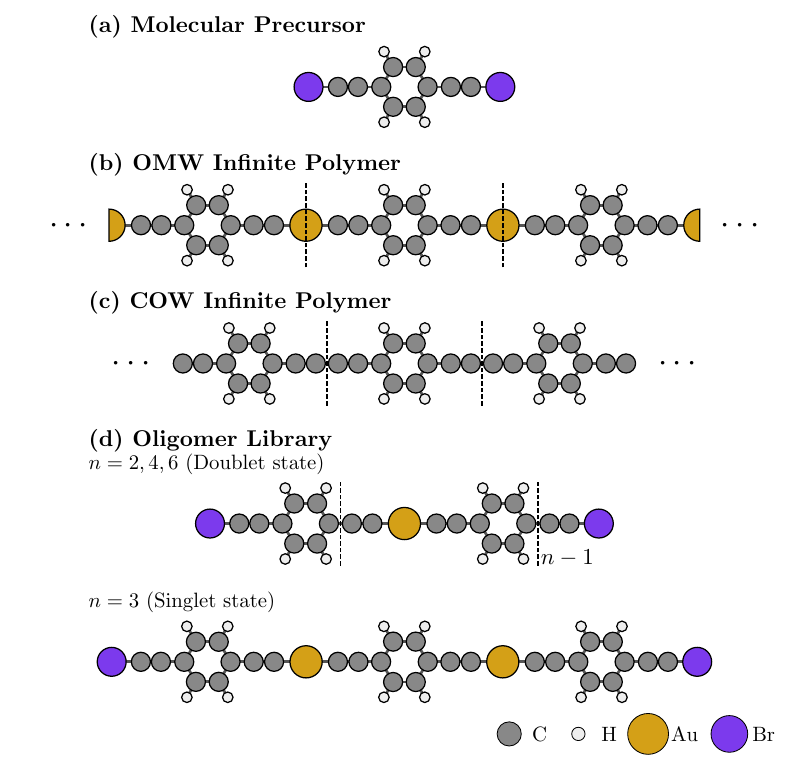}
\caption{Optimized PBE+MBD geometries. (a) Molecular precursor 1,4-bis(bromoethynyl)benzene. (b) Organometallic wire (OMW) infinite polymer with $\mathrm{C{-}Au{-}C}$ bridges. (c) Covalent organic wire (COW) as an infinite polymer with continuous diacetylenic linkages. (d) Finite oligomer library, $n = 2, 4, 6$ doublet radicals and $n = 3$ closed-shell singlet. Carbon atoms are indicated in gray, hydrogen in white, gold in gold, and bromine in purple, with dashed lines marking primitive-cell or formula-unit boundaries.}
\label{fig:geometry}
\end{figure}

The chemical identity of the linkage between consecutive phenylene units defines the two phases of the free-standing $\gamma$-graphdiyne molecular wire. In the covalent organic wire (COW), the linkage is a diacetylenic $\mathrm{-C{\equiv}C{-}C{\equiv}C-}$ unit that preserves an uninterrupted $\pi$-conjugation pathway along the chain. In the organometallic wire (OMW), the central single bond of the diyne is replaced by a $\mathrm{C{-}Au{-}C}$ bridge that interrupts $\pi$-conjugation and introduces a metal center carrying eleven valence electrons per formula unit. The relaxed PBE+MBD geometries of both polymers, together with the molecular precursor and the finite-oligomer library used in the length-convergence analysis, are shown in Fig.~\ref{fig:geometry}.

The molecular precursor of the on-surface synthesis (Fig.~\ref{fig:geometry}a) is 1,4-bis(bromoethynyl)benzene, a closed-shell aromatic carrying two terminal $\mathrm{-C{\equiv}C{-}Br}$ groups whose thermal dehalogenation on Au(100) releases the bromine atoms and exposes the reactive sites that drive the polymerization. The relaxed geometry of the precursor is reported here as the molecular reference against which the two polymeric phases are compared.

The organometallic wire (Fig.~\ref{fig:geometry}b) relaxes to a periodicity of $c = 11.9944$~\AA{} per $\mathrm{C_{10}H_{4}Au}$ formula unit. Replacing the central single bond of the diyne with the $\mathrm{C{-}Au{-}C}$ bridge produces two structural fingerprints that identify the change in electronic character. The $\mathrm{C{\equiv}C}$ triple bond elongates to $1.245$~\AA{}, a $1.3\%$ stretching relative to COW, while the $\mathrm{C(sp^{2}){-}C(sp)}$ bond contracts to $1.391$~\AA{}, a $1.1\%$ shortening relative to COW. Both effects are the geometric signature of partial cumulenization of the chain, in which the $\pi$ density redistributes from the localized triple bonds toward the ring--acetylene junction and the backbone approaches the bond-equalized cumulenic limit. The microscopic origin is the back-donation $\mathrm{Au}(5d) \to \pi^{*}(\mathrm{C{\equiv}C})$, which weakens the triple bond while reinforcing conjugation across the $\mathrm{C(sp^{2}){-}C(sp)}$ junction. The relaxed $\mathrm{Au{-}C}$ bond of $1.946$~\AA{} lies in the range typical of organogold compounds and is compatible with a covalent rather than ionic description of the metal-carbon linkage, supported by the small Hirshfeld charges $q_\mathrm{Au} = +0.146~e$ on the metal and $q_\mathrm{C(sp)} = -0.151~e$ on the adjacent carbons (Table~S4 of the Supporting Information).

The covalent organic wire (Fig.~\ref{fig:geometry}c) relaxes to a periodicity of $c = 9.4464$~\AA{} per primitive cell containing 14 atoms. The value reproduces the experimental periodicity of approximately $0.95$~nm reported by Cartoceti \textit{et al.}~\cite{Cartoceti2026} on Au(100) to within $-0.5\%$, an agreement that lies well inside the expected accuracy of the PBE+MBD treatment of $\pi$-conjugated organic polymers. The relaxed bond pattern exhibits the alternating single--triple sequence characteristic of polyynic conjugation, with $\mathrm{C{\equiv}C}$ triple bonds of $1.229$~\AA{}, $\mathrm{C(sp){-}C(sp)}$ single bonds of $1.346$~\AA{} inside the diyne unit, $\mathrm{C(sp^{2}){-}C(sp)}$ bonds of $1.406$~\AA{} at the ring--acetylene junction, and aromatic $\mathrm{C(sp^{2}){-}C(sp^{2})}$ bonds of $1.415$~\AA{} within the phenylene ring. The bond-length alternation $\mathrm{BLA} = r_\mathrm{single} - r_\mathrm{triple}$ along the diyne segment is $0.117$~\AA, in agreement with the values reported for $\pi$-conjugated polyynic systems of comparable backbone length and supporting the strongly polyynic character of the COW chain. The dispersion energies recovered by the MBD correction amount to $-0.394$~eV per primitive cell for COW and $-0.425$~eV for OMW, a modest contribution that does not modify the relative ordering of bond lengths but is considered throughout the present work for consistency with the strain energetics discussed below.

The OMW oligomer library (Fig.~\ref{fig:geometry}d) spans $n = 2, 3, 4, 6$ with bromine termination, reproducing the finite-domain structures observed by scanning tunneling microscopy after the room-temperature dehalogenation step. The number of Au atoms per oligomer is $n - 1$, so that the total valence-electron count is odd for $n = 2, 4, 6$ and even only for $n = 3$. The odd-electron cases form a series of $\pi$-radical doublets whose unpaired electron is distributed over the carbon backbone with a small admixture on Au, while $n = 3$ provides the unique closed-shell singlet reference of the series. The HSE06 Kohn-Sham eigenvalues of the four oligomers and of the precursor are reported in Figure~S8 of the Supporting Information and are used in the size-convergence analysis discussed below. Equilibrium structural descriptors of both polymers are collected in Table~\ref{tab:eq_props}.

\begin{table}[h!]
\centering
\caption{Equilibrium structural and electronic descriptors of the free-standing covalent (COW) and organometallic (OMW) $\gamma$-graphdiyne molecular wires from PBE+MBD relaxations. The bond-length alternation $\mathrm{BLA} = r_\mathrm{single} - r_\mathrm{triple}$ is taken along the diyne segment of COW. $E^\mathrm{MBD}_\mathrm{vdW}$ is the many-body dispersion contribution to the total energy at equilibrium, evaluated within the MBD@rsSCS scheme~\cite{Tkatchenko2012,Ambrosetti2014}. Hirshfeld charges $q_\mathrm{Au}$ are reported on the Au site of the OMW.}
\label{tab:eq_props}
\begin{tabular}{lcc}
\hline\hline
Property & COW & OMW \\
\hline
Atoms per primitive cell & 14 & 15 \\
Valence electrons per cell & 44 & 55 \\
Lattice parameter $c$ (\AA) & $9.4464$ & $11.9944$ \\
$r_\mathrm{C(sp^{2}){-}C(sp^{2})}$ aromatic, long/short (\AA) & $1.415\,/\,1.383$ & $1.427\,/\,1.374$ \\
$r_\mathrm{C(sp^{2}){-}C(sp)}$ (\AA) & $1.406$ & $1.391$ \\
$r_\mathrm{C{\equiv}C}$ triple (\AA) & $1.229$ & $1.245$ \\
$r_\mathrm{C(sp){-}C(sp)}$ diyne (\AA) & $1.346$ & -- \\
$r_\mathrm{Au{-}C(sp)}$ (\AA) & -- & $1.946$ \\
$\mathrm{BLA} = r_\mathrm{single} - r_\mathrm{triple}$ (\AA) & $0.117$ & -- \\
$E^\mathrm{MBD}_\mathrm{vdW}$ (eV/cell) & $-0.394$ & $-0.425$ \\
Hirshfeld $q_\mathrm{Au}$ ($e$) & -- & $+0.146$ \\
\hline\hline
\end{tabular}
\end{table}

\begin{figure}[htb!]
\centering
\includegraphics[width=\linewidth]{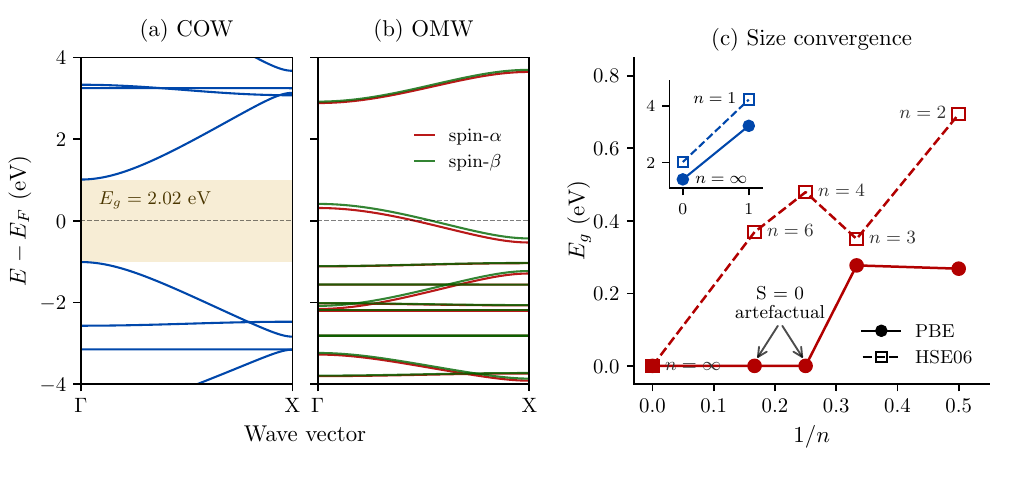}
\caption{HSE06 electronic structure of the free-standing $\gamma$-graphdiyne molecular wires on the relaxed PBE+MBD geometries. (a) COW polymer along $\Gamma \to \mathrm{X}$, with the direct gap of $E_g = 2.02$~eV at $\Gamma$ shaded. (b) OMW polymer in the spin-collinear ferromagnetic configuration, with spin-$\alpha$ (red) and spin-$\beta$ (dark green) channels overlaid. (c) Size convergence of the gap $E_g(1/n)$, PBE+MBD (filled circles, solid lines) and HSE06 (open squares, dashed lines), with the inset showing the COW endpoints at $n = 1$ and $1/n = 0$. Lines are guides to the eye.}
\label{fig:electronic}
\end{figure}

Building on the equilibrium structural picture, we turn now to the electronic dispersion of both phases. The structural distinction between the polyynic COW and the partially cumulenized OMW has a direct counterpart in the electronic structure, which is best resolved at the hybrid-functional level because GGA functionals systematically underestimate electronic band gaps in $\pi$-conjugated polymers and fail to localize unpaired electrons on multi-metal sites. Figure~\ref{fig:electronic} presents the HSE06 electronic band structures of both polymers along $\Gamma \to \mathrm{X}$ on the relaxed PBE+MBD geometries, together with the size convergence of the electronic band gap toward the polymer limit.

The covalent organic wire (Fig.~\ref{fig:electronic}a) is a one-dimensional semiconductor with a direct electronic band gap of $E_g^\mathrm{HSE06} = 2.02$~eV at $\Gamma$, originating from the alternating single--triple bond pattern of the diacetylenic chain. The corresponding PBE+MBD value of $E_g^\mathrm{PBE} = 1.41$~eV lies approximately $0.6$~eV below the HSE06 result, a systematic underestimation by GGA that is well documented for $\pi$-conjugated 1D polymers and is largely resolved by the $25\%$ Hartree--Fock admixture of HSE06~\cite{Krukau2006}. The valence-band maximum and conduction-band minimum coincide in $k$-space at $\Gamma$, placing COW in the direct-gap regime relevant to optical transitions, with an electronic structure that aligns with the $\pi$-conjugated polymer family of poly($p$-phenylene-butadiynylene).

The organometallic wire (Fig.~\ref{fig:electronic}b) presents a qualitatively different picture. The dense band-structure path along $\Gamma \to \mathrm{X}$, sampled at $101$ wave vectors, shows that the OMW polymer is metallic at the HSE06 level, with the spin-$\alpha$ and spin-$\beta$ channels of the collinear ferromagnetic solution crossing the Fermi level at distinct wave vectors of $k_z \approx 0.23 \cdot 2\pi/c$ and $k_z \approx 0.27 \cdot 2\pi/c$ respectively. The crossing is confirmed by the fractional occupation $f \approx 0.5$ of the two bands at their respective Fermi-crossing $k$-points, the analytical signature of a band that traverses $E_F$ under Gaussian smearing. The two crossings are separated in $k$ by approximately $0.04 \cdot 2\pi/c$ but are uncoupled by the spin-collinear treatment and do not open a gap, so that the spectral weight at $E_F$ is finite from both spin channels. The complete set of HSE06 band structures of the four ground-state probes investigated in this work is reported in Figure~S5 of the Supporting Information. The two phases, therefore, differ qualitatively in their electronic ground state, with COW a wide-gap semiconductor and OMW a one-dimensional metal whose effective conjugation is interrupted by the $\mathrm{C{-}Au{-}C}$ bridge.

The size convergence of the electronic band gap (Fig.~\ref{fig:electronic}c) probes how this contrast emerges as the chain length grows from the molecular to the polymeric limit. For COW, only the molecular reference at $n = 1$ (1,4-diethynylbenzene, $E_g^\mathrm{HSE06} = 4.22$~eV) and the polymer ($E_g^\mathrm{HSE06} = 2.02$~eV) are reported here, while intermediate-length COW oligomers are an obvious extension for future work. The two-point evolution is consistent with the documented behavior of polyyne-type chains, in which the gap decreases smoothly with chain length and approaches the polymeric limit by $n \gtrsim 10$~\cite{Yang2008,Cinquanta2011}; the two present values bracket that progression at its molecular and periodic extremes. For OMW, the denser series ($n = 2, 3, 4, 6$ plus polymer) reveals a visibly non-monotonic envelope $\{0.69, 0.35, 0.48, 0.37, 0\}$~eV at HSE06 that encodes the interplay between three effects, namely Au-parity alternation between odd-$n$ closed-shell singlets and even-$n$ doublet radicals, the systematic failure of PBE+MBD in describing the multi-Au radicals. The progressive delocalization of the radical along the chain that, in the polymeric limit, dissolves the localized singly occupied molecular orbital (SOMO) into a Bloch state crossing the Fermi level and yields the metallic dispersion discussed above. For the open-shell oligomers ($n = 2, 4, 6$) the electronic band gap reported in Fig.~\ref{fig:electronic}c is the fundamental gap of the doublet, defined as the lowest unoccupied molecular orbital (LUMO) minus the highest occupied molecular orbital (HOMO) taken across the two spin channels, $E_g = \min_\sigma\,\mathrm{LUMO}_\sigma - \max_{\sigma'}\,\mathrm{HOMO}_{\sigma'}$, rather than within a single channel. The smallest oligomer ($n = 2$) is the most asymmetric case, with the single $\mathrm{Au}$ atom hosting a strongly localized SOMO that splits the spin-$\alpha$ and spin-$\beta$ manifolds by several eV and reduces the fundamental gap to the intra-spin-$\beta$ separation, while for $n = 4$ and $n = 6$ the radical delocalizes over the multi-Au backbone and the two spin channels approach each other to within a few tens of meV. The spin-resolved frontier orbitals that produce these fundamental gaps are reported in Figure~S8 of the Supporting Information.

The role of the functional becomes explicit upon comparing PBE+MBD and HSE06 for the same oligomers. At the PBE+MBD level, the alternating $\pm 1$ initial moment on the Au sublattice does not survive the SCF cycle for $n = 4$ and $n = 6$, and the system collapses to an artefactual non-magnetic configuration with $S = 0$ and vanishing gap. This collapse is the classical signature of over-delocalization in GGA, in which the unpaired electron spreads symmetrically across the Au sublattice with fractional $\alpha/\beta$ occupations, an arrangement preferred by GGA because of its uncorrected self-interaction error~\cite{Cohen2008,MoriSanchez2008}. The HSE06 functional, with its $25\%$ exact-exchange admixture, penalizes the spurious symmetric solution and restores the localized doublet at $n = 4$ ($E_g = 0.48$~eV) and $n = 6$ ($E_g = 0.37$~eV), confirming that the GGA result is methodological in origin rather than physical. The dip at $n = 3$ in HSE06 is in turn a real consequence of the closed-shell character, since with no unpaired electron available, no spin-symmetry breaking can open the gap, and the value $E_g^\mathrm{HSE06}(n = 3) = 0.35$~eV reflects the bare $\pi$-conjugation of the closed-shell oligomer.

The Kuhn-type model $E_g(n) = a + b/(n + c)$ that is standard for polyene and polyyne chains~\cite{Casari2016} assumes a monotonic length dependence governed by a single conjugation parameter, an assumption that the OMW series violates because of the parity-dependent spin state, so that no global three-parameter Kuhn fit can describe the data and we therefore present the points connected only by guides to the eye in Fig.~\ref{fig:electronic}c. The qualitative observation that the gap evolution of OMW is governed by spin localization rather than by uniform $\pi$-conjugation is itself the central result of this length-dependent analysis, and it underscores the necessity of hybrid functionals for any sensible treatment of the organometallic phase.

\begin{figure}[t!]
\centering
\includegraphics[width=\linewidth]{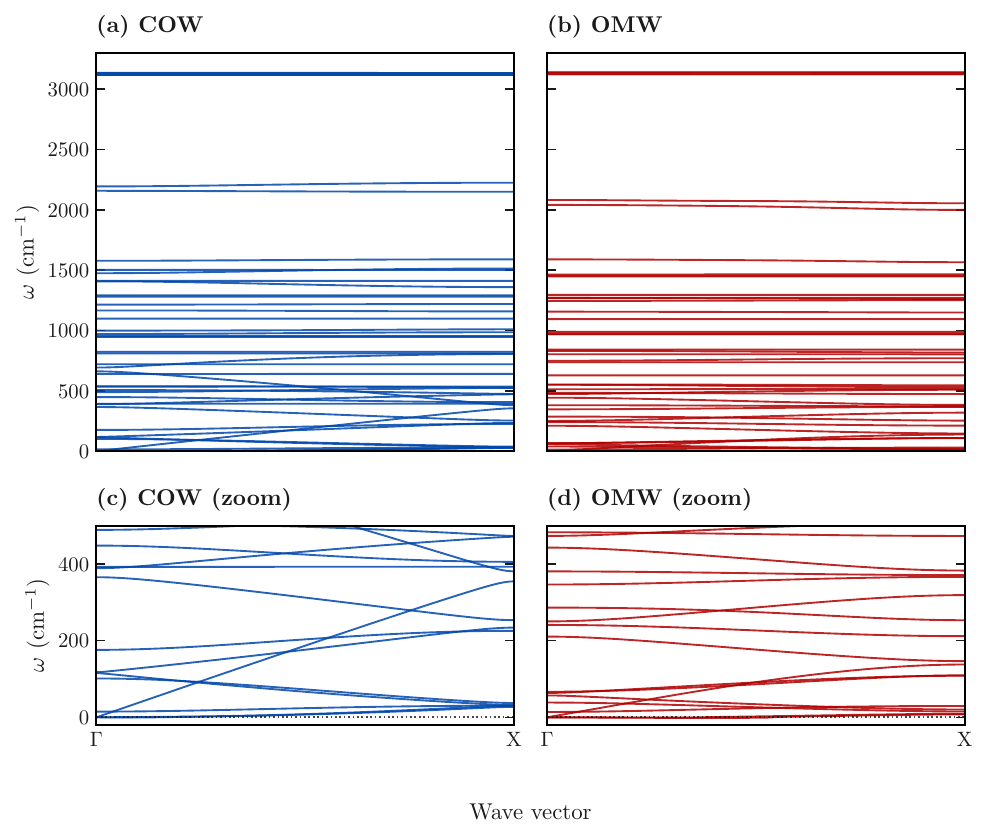}
\caption{Phonon dispersion of the free-standing $\gamma$-graphdiyne molecular wires along $\Gamma \to \mathrm{X}$, computed by finite displacements in Phonopy on a $1 \times 1 \times 6$ supercell of the PBE+MBD-relaxed primitive cell. (a, b) Full range. (c, d) Zoomed view of the acoustic region. The dashed line marks $\omega = 0$, the threshold for dynamical instability.}
\label{fig:phonons}
\end{figure}

The structural and electronic distinctions between the two phases have lattice-dynamical counterparts that probe both the bonding stiffness and the absolute stability of the free-standing wires. Figure~\ref{fig:phonons} presents the full phonon dispersion of both polymers along $\Gamma \to \mathrm{X}$ at the PBE+MBD level, in the upper row, together with a zoomed view of the acoustic region in the lower row that addresses the stability question directly.

The full dispersions in Fig.~\ref{fig:phonons}(a, b) extend from the acoustic branches at $\omega = 0$ to the C-H stretching manifold near $3120$~cm$^{-1}$. The two polymers share the same aromatic-phenylene backbone, and the corresponding fingerprints (the C-H stretches near $3120$~cm$^{-1}$, the G-band-like aromatic ring stretch near $1580$~cm$^{-1}$, and the in-plane ring deformations spanning $700$--$1500$~cm$^{-1}$) appear at essentially the same frequencies in both phases. The qualitative differences are confined to two regions. The first is the $\mathrm{C}{\equiv}\mathrm{C}$ triple-bond doublet, which sits at $2157.6$~cm$^{-1}$ and $2193.9$~cm$^{-1}$ in COW and at $2040.8$~cm$^{-1}$ and $2081.9$~cm$^{-1}$ in OMW, corresponding to a uniform downshift of approximately $-115$~cm$^{-1}$ on insertion of the $\mathrm{C{-}Au{-}C}$ bridge. The OMW value of $2081.9$~cm$^{-1}$ reproduces the Raman peak observed by Cartoceti \textit{et al.}~\cite{Cartoceti2026} near $2080$~cm$^{-1}$ on Au(100) to within $5$~cm$^{-1}$, evidencing the agreement of the present free-standing calculation with the experimental fingerprint of the organometallic phase. The $-115$~cm$^{-1}$ shift relative to the diacetylenic COW provides, in turn, a quantitative interpretation of the chemical effect of the gold bridge that the experiment alone cannot isolate, since the on-surface Raman signal mixes the intrinsic chain response with the surface-induced renormalization of the same modes. The second qualitative difference is the region below $400$~cm$^{-1}$, where the OMW dispersion in Fig.~\ref{fig:phonons}(b) shows additional optical branches that are absent in the COW and that involve substantial Au displacement. These low-frequency branches, together with the high-frequency C-H stretches, the C$_{sp^{2}}$ aromatic ring modes, and the C$_{sp}$ alkynyl stretches, constitute the four characteristic regimes of the vibrational spectrum, with the quantitative species-resolved assignment of every mode left to the next subsection.

A central question for the free-standing wire, beyond its surface-bound experimental observation~\cite{Cartoceti2026}, is whether either phase is dynamically stable without the supporting metal substrate. The zoom panels in Fig.~\ref{fig:phonons}(c, d) address this question directly. The four acoustic branches of each polymer (one longitudinal, two transverse, and one torsional around the chain axis, characteristic of a one-dimensional polymer in three-dimensional space) reach $\omega = 0$ at $\Gamma$ within numerical noise. In COW, the entire $\Gamma \to \mathrm{X}$ path remains non-negative, with the largest span across the supercell convergence tests below $0.4$~cm$^{-1}$, well within the finite-displacement limit of the harmonic approximation. In OMW, the soft torsional acoustic branch shows imaginary values of up to $\approx 1.7$~cm$^{-1}$ over part of the $\Gamma \to \mathrm{X}$ path, of the order of the noise limit expected for a torsional mode of a one-dimensional polymer in three-dimensional vacuum at this displacement amplitude. No genuine soft branch capable of driving a Peierls-like dimerization or a structural symmetry breaking is observed in either phase, and the free-standing OMW polymer is therefore dynamically stable in the harmonic sense despite its metallic character. This result is itself a contribution of the present work, since the phonon information reported in~\cite{Cartoceti2026} was restricted to the zone center and to substrate-decoupled fragments and cannot separate the intrinsic stability of the isolated chain from the stabilizing effect of the Au(100) substrate.

Integration of the harmonic spectrum on a $1 \times 1 \times 60$ phonon mesh yields zero-point energies of $256.30$~kJ/mol and $253.91$~kJ/mol per primitive cell for COW and OMW, respectively, a small difference of approximately $2.4$~kJ/mol per cell that reflects the substitution of one carbon atom by a heavier gold atom in the OMW formula unit. On a per-atom basis, the values are $0.190$~eV/atom for COW and $0.176$~eV/atom for OMW, with the smaller OMW figure tracking the lower vibrational frequencies of the Au-containing modes. The temperature dependence of the harmonic free energy $F_\mathrm{vib}(T)$, the heat capacity at constant volume $C_V(T)$, and the vibrational entropy $S(T)$ obtained from the same integration carry complementary physical information about the two phases. At $T = 300$~K the heat capacities are $142.6$~J\,mol$^{-1}$\,K$^{-1}$ for COW and $162.7$~J\,mol$^{-1}$\,K$^{-1}$ for OMW, corresponding respectively to $41\%$ and $44\%$ of the classical Dulong--Petit limit $3Nk_B$, so that both polymers are well within the Einstein-like regime dominated by the high-frequency aromatic ring and $\mathrm{C}{\equiv}\mathrm{C}$ stretches that remain frozen at room temperature. The larger $C_V$ of OMW originates from the manifold of Au-projected optical branches below $400$~cm$^{-1}$ that approach the equipartition limit at ambient conditions, whereas the corresponding low-frequency region of COW carries fewer modes per cell. The same low-frequency manifold is responsible for the larger vibrational entropy of OMW, $S^\mathrm{OMW}(300~\mathrm{K}) = 237.2$~J\,mol$^{-1}$\,K$^{-1}$ versus $S^\mathrm{COW}(300~\mathrm{K}) = 202.8$~J\,mol$^{-1}$\,K$^{-1}$, an entropic bias of $\Delta S = +34.4$~J\,mol$^{-1}$\,K$^{-1}$ at room temperature that translates into a $T \Delta S$ stabilization of $+10.3$~kJ/mol favoring the organometallic phase. The total vibrational free energy at $300$~K therefore favors OMW over COW by $\Delta F_\mathrm{vib}(300~\mathrm{K}) = -8.1$~kJ/mol per cell, with the entropic gain exceeding the zero-point penalty and consistent with the room-temperature stability of the OMW intermediate observed in the on-surface synthesis~\cite{Cartoceti2026}.

\begin{figure}[t!]
\centering
\includegraphics[width=\linewidth]{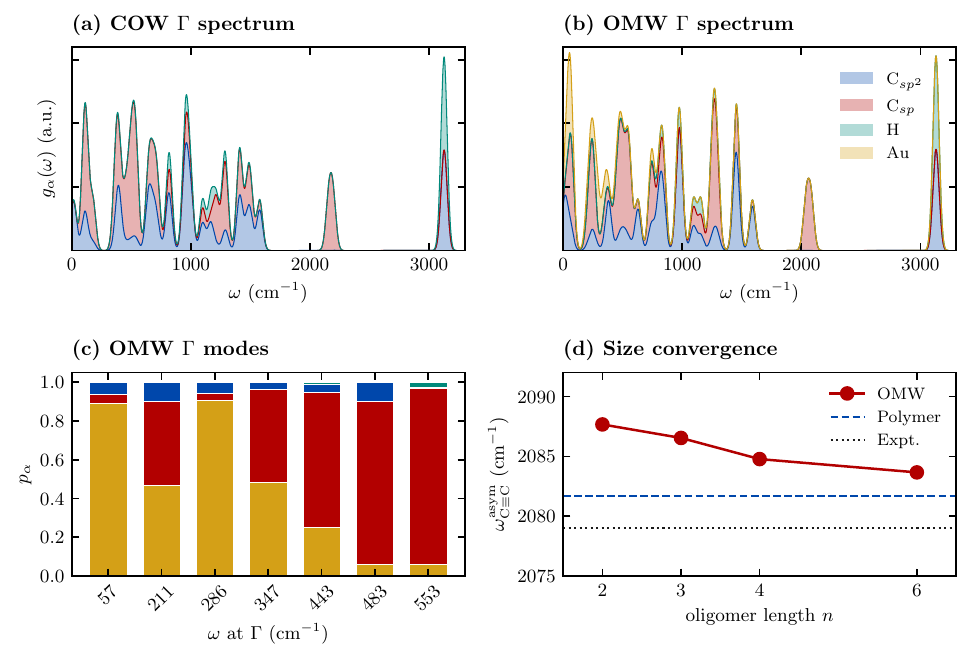}
\caption{Species-resolved vibrational analysis of the free-standing $\gamma$-graphdiyne molecular wires. (a) COW $\Gamma$-point spectrum decomposed into $\mathrm{C}_{sp^{2}}$, $\mathrm{C}_{sp}$, H, and Au contributions. (b) OMW $\Gamma$-point spectrum in the same decomposition. (c) OMW $\Gamma$-point modes with $p_\mathrm{Au} > 0.05$ in the window $50$--$700$~cm$^{-1}$, stacked by atomic species. (d) Convergence of the highest antisymmetric $\mathrm{C}{\equiv}\mathrm{C}$ frequency along the OMW oligomer series toward the polymeric limit (dashed line) and the experimental Raman position of Cartoceti \textit{et al.}\ (dotted line).}
\label{fig:fingerprints}
\end{figure}

The full phonon dispersion shows that the two free-standing phases are dynamically stable and identifies the $\mathrm{C}{\equiv}\mathrm{C}$ doublet as the spectral region where they differ most strongly. However, this does not address how each atomic species contributes to the eigenvectors of the modes that an experimental Raman measurement would isolate. The mode-character decomposition $p_\alpha(\nu, q)$ defined in Eq.~\ref{eq:participation} answers this question by projecting every eigenvector onto contributions from gold, $sp$ carbon, $sp^{2}$ carbon, hydrogen, and bromine, with the carbon topological labels fixed by the connectivity criterion of the Methodology and stable under axial strain. Figure~\ref{fig:fingerprints} presents the resulting species-resolved spectra for both polymers together with the bar diagram of all OMW modes carrying $p_\mathrm{Au} > 0.05$ and the size convergence of the antisymmetric $\mathrm{C}{\equiv}\mathrm{C}$ mode along the finite-oligomer series.

The two stacked partial densities of vibrational states in Fig.~\ref{fig:fingerprints}(a, b) reveal that both polymers share the same four canonical spectral regions inherited from the aromatic-phenylene backbone, namely the C-H stretches near $3120$~cm$^{-1}$ that are essentially equal mixtures of $\mathrm{C}_{sp^{2}}$ and H character, the G-band-like aromatic stretch at $1577.6$~cm$^{-1}$ in COW and $1589.5$~cm$^{-1}$ in OMW dominated by $\mathrm{C}_{sp^{2}}$, the in-plane ring deformations spanning $700$--$1500$~cm$^{-1}$, and the high-frequency $\mathrm{C}{\equiv}\mathrm{C}$ doublet near $2160$ and $2080$~cm$^{-1}$ respectively. The decomposition isolates two qualitative differences between the two polymers. The first concerns the $\mathrm{C}{\equiv}\mathrm{C}$ doublet itself, where both the symmetric and the antisymmetric components are $99$--$100\%$ $\mathrm{C}_{sp}$ in either phase, with no detectable Au participation in the eigenvector of OMW. The $-115$~cm$^{-1}$ downshift of the doublet from COW to OMW is therefore not the result of vibrational hybridization of the C and Au degrees of freedom, but reflects exclusively the electronic softening of the triple-bond force constant produced by the back-donation $\mathrm{Au}(5d) \to \pi^{*}(\mathrm{C}{\equiv}\mathrm{C})$ identified at the geometric level in the equilibrium-geometry analysis above. The second difference is the ECC mode at $1407.6$~cm$^{-1}$ in COW, a collective stretch of the single bonds of the $sp$-carbon backbone with $87\%$ $\mathrm{C}_{sp}$ and $13\%$ $\mathrm{C}_{sp^{2}}$ character, which has no counterpart in OMW because the corresponding single bond has been replaced by the $\mathrm{C{-}Au{-}C}$ bridge. The ECC mode is therefore a vibrational fingerprint specific to the polyynic COW phase, and its absence is a diagnostic of OMW formation.

The bar diagram in Fig.~\ref{fig:fingerprints}(c) collects all OMW modes for which $p_\mathrm{Au}$ exceeds $0.05$ within the window $50$--$700$~cm$^{-1}$ and decomposes each mode into the four atomic species. Two modes at $57.0$~cm$^{-1}$ and $286.2$~cm$^{-1}$ are essentially pure Au with $p_\mathrm{Au} = 0.89$ and $0.91$ respectively, corresponding to a translational wagging of the metal site and to a longitudinal Au displacement along the chain axis. Two mixed modes at $210.7$~cm$^{-1}$ and $346.8$~cm$^{-1}$ have nearly balanced $\mathrm{Au}$ and $\mathrm{C}_{sp}$ participation and correspond to $\mathrm{Au{-}C}$ stretching and bending vibrations of the rigid metal-carbon link. The entire Au-projected manifold sits below $400$~cm$^{-1}$, a region that is empty of organic-backbone activity in COW and that therefore constitutes a vibrational fingerprint specific to the organometallic phase. A Raman peak observed below $400$~cm$^{-1}$ on an on-surface sample is, accordingly, a direct signature of OMW domains, and the pair $57$ and $286$~cm$^{-1}$ at nearly pure Au character is particularly suited to low-frequency Raman detection of residual organometallic regions during the OMW-to-COW transition.

The convergence panel in Fig.~\ref{fig:fingerprints}(d) assesses the legitimacy of comparing the polymeric calculation with on-surface Raman measurements that probe a distribution of finite-length domains. The maximum frequency of the antisymmetric $\mathrm{C}{\equiv}\mathrm{C}$ stretch of the internal Au-bound triple bonds saturates rapidly with the oligomer length $n$, with a residual of $5.8$~cm$^{-1}$ between $n = 2$ and the polymer and only $1.8$~cm$^{-1}$ between $n = 6$ and the polymer. Both values are well below the intrinsic accuracy of PBE+MBD for Raman-active modes of $\pi$-conjugated polymers, of the order of $10$--$30$~cm$^{-1}$, so that any oligomer with at least two internal Au bridges already reproduces the polymeric Raman fingerprint within the relevant uncertainty. This rapid convergence justifies quantitatively the comparison between the polymeric calculation reported here and the experimental Raman signal of Cartoceti \textit{et al.}~\cite{Cartoceti2026}, which arises from on-surface domains of distributed but short lengths.

\begin{figure}[t!]
\centering
\includegraphics[width=\linewidth]{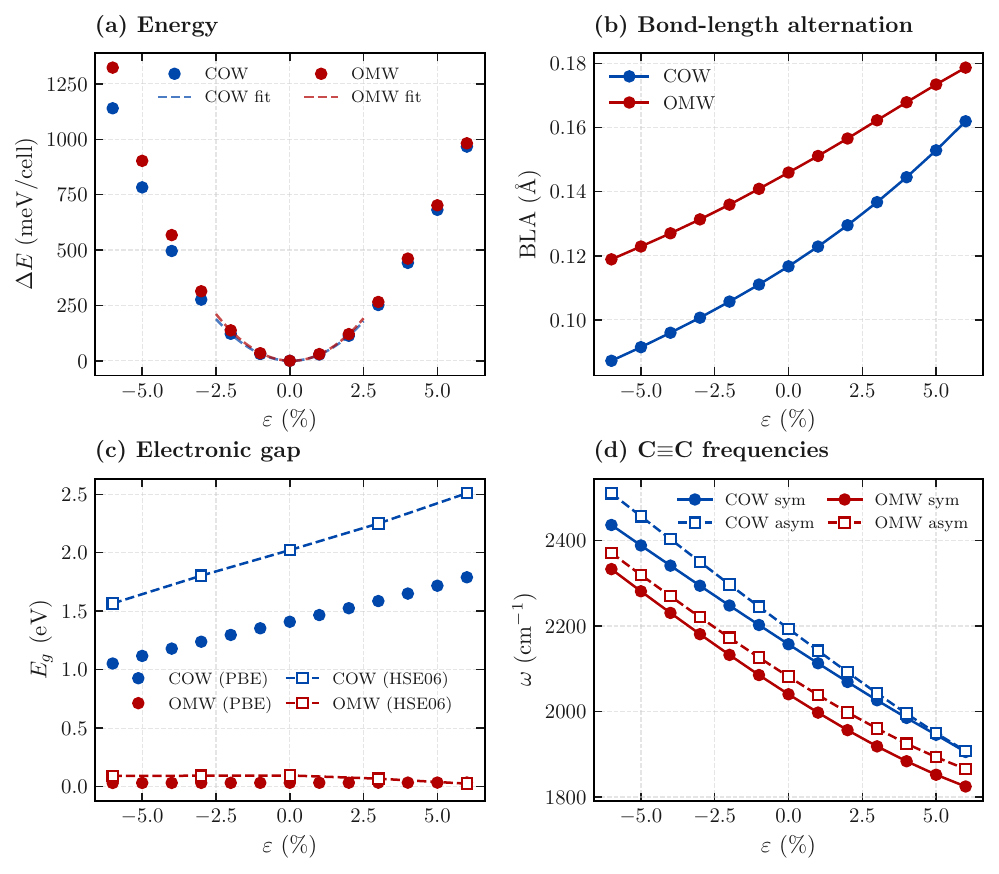}
\caption{Response of the free-standing $\gamma$-graphdiyne molecular wires to axial strain $\varepsilon \in [-6, +6]\%$. (a) Total energy per primitive unit cell relative to the equilibrium configuration, with quadratic fits in the linear regime $|\varepsilon| \leq 2\%$ used to extract the one-dimensional Young's modulus values. (b) Bond-length alternation for COW. (c) Electronic band gap values at PBE+MBD (filled circles) and HSE06 (open squares, dashed lines). (d) Symmetric (circles) and antisymmetric (squares, dashed lines) $\mathrm{C}{\equiv}\mathrm{C}$ frequencies. COW in blue, OMW in red.
}
\label{fig:strain}
\end{figure}

The chemistry of the $\mathrm{C{-}Au{-}C}$ bridge identified at the structural and vibrational levels raises a complementary question about the mechanical and electronic response of each phase to axial deformation, which is the experimentally relevant figure of merit for one-dimensional carbon wires both under thermal fluctuation and under tensile loading by a scanning probe. The three-layer protocol described in the Methodology samples the relaxed polymer at thirteen strain values $\varepsilon \in [-6, +6]\%$ at the PBE+MBD level, with HSE06 single-points on the subset $\{-6, -3, 0, +3, +6\}\%$ and $\Gamma$-only Phonopy calculations at every strain. The complete energy and lattice dataset of the PBE+MBD strain protocol is reported in Table~S3 of the Supporting Information, and the HSE06 band structures of both polymers along the sampled strain range are reported in Figures~S6 and~S7. Figure~\ref{fig:strain} reports the resulting curves for the energy, the bond-length alternation, the electronic gap, and the symmetric and antisymmetric $\mathrm{C}{\equiv}\mathrm{C}$ frequencies.

The total-energy response in Fig.~\ref{fig:strain}(a) is parabolic in the linear regime $|\varepsilon| \leq 2\%$ for both polymers, with the quadratic fit yielding one-dimensional Young's moduli of $Y^\mathrm{COW}_\mathrm{1D} = 995$~nN and $Y^\mathrm{OMW}_\mathrm{1D} = 861$~nN, a reduction of $13.5\%$ on insertion of the $\mathrm{C{-}Au{-}C}$ bridge. The structural distribution of the imposed deformation is qualitatively distinct between the two phases. In COW, a tensile strain of $+6\%$ is shared among the three principal bonds along the chain, namely the $\mathrm{C}{\equiv}\mathrm{C}$ triples elongated by $+2.7\%$, the internal $\mathrm{C(sp){-}C(sp)}$ single bond of the diyne by $+5.8\%$, and the $\mathrm{C(sp^{2}){-}C(sp)}$ junction by $+7.0\%$. In OMW, the same $+6\%$ macroscopic strain is concentrated on the $\mathrm{Au{-}C}$ bond, which stretches by $+9.2\%$ while the $\mathrm{C}{\equiv}\mathrm{C}$ triples and the $\mathrm{C(sp^{2}){-}C(sp)}$ junctions absorb only $+2.3\%$ and $+4.4\%$ respectively. The $\mathrm{Au{-}C}$ link accordingly takes up more than half of the total axial deformation, identifying it as the softest mechanical degree of freedom of the organometallic chain and providing the structural origin of the reduced Young's modulus.

The bond-length alternation reported in Fig.~\ref{fig:strain}(b) traces these geometric trends. In COW, the BLA $= r_\mathrm{single} - r_\mathrm{triple}$ along the diyne segment evolves linearly from $\mathrm{BLA}(\varepsilon = -6\%) \approx 0.087$~\AA{} to $\mathrm{BLA}(\varepsilon = +6\%) \approx 0.162$~\AA, with the equilibrium value of $0.117$~\AA{} characteristic of strong polyynic conjugation. In OMW, the analogous quantity defined between the $\mathrm{C}{\equiv}\mathrm{C}$ triple bond and the adjacent $\mathrm{C(sp^{2}){-}C(sp)}$ junction is buffered by the $\mathrm{Au{-}C}$ bond and varies over a comparable range, but the deformation absorbed by the Au-carbon link decouples the alternation from the chain-axis strain, so that the partial cumulenization of the OMW backbone is preserved over a wider strain window than in the purely covalent phase.

The electronic-gap response shown in Fig.~\ref{fig:strain}(c) separates the two phases by their qualitative behavior. In COW, both PBE+MBD and HSE06 reproduce the classical Peierls-type coupling of a polyynic chain, with the gap opening under tension and closing under compression. The PBE+MBD slope is $\partial E_g / \partial \varepsilon \approx +6.2$~eV/strain, and the HSE06 slope amplifies to approximately $+8$~eV/strain, with the $25\%$ Hartree--Fock admixture penalizing the over-delocalized PBE solution and reinforcing the bond-length alternation that controls the gap. In OMW, the system remains metallic across the entire $\varepsilon \in [-6, +6]\%$ range at both PBE+MBD and HSE06 levels, with the spin-$\alpha$ and spin-$\beta$ Fermi crossings of the collinear ferromagnetic solution shifting only marginally in $k$ along $\Gamma \to \mathrm{X}$ as the lattice parameter is varied. The strain-electronic response of OMW is therefore qualitatively distinct from that of COW, with the structural and vibrational degrees of freedom carrying the diagnostic signal of axial deformation. The contrast between phases is significant, with the polyynic phase showing Peierls coupling between the BLA and the gap, and the organometallic phase preserving its metallic character under axial strain.

The mode-resolved Gr\"uneisen parameters extracted from the $\mathrm{C}{\equiv}\mathrm{C}$ frequencies in Fig.~\ref{fig:strain}(d) yield $\gamma^\mathrm{COW}_\mathrm{sym} = 2.06$ and $\gamma^\mathrm{COW}_\mathrm{asym} = 2.31$ for the covalent phase and $\gamma^\mathrm{OMW}_\mathrm{sym} = 2.08$ and $\gamma^\mathrm{OMW}_\mathrm{asym} = 2.03$ for the organometallic phase. The absolute magnitudes lie within the range typical of $sp$-carbon wires and slightly above the values reported for polyenes, in accordance with the strong anharmonicity of triple-bond stretches under axial strain. The qualitative difference between phases is related to the relative magnitudes of $\gamma_\mathrm{sym}$ and $\gamma_\mathrm{asym}$. In COW, the antisymmetric component is more sensitive to strain than the symmetric one because the two triple bonds are coupled through the central single bond of the diyne, which experiences a large fraction of the strain. In OMW, the same components have nearly identical Gr\"uneisen coefficients because the two triple bonds are decoupled by the soft $\mathrm{Au{-}C}$ link, which stands the strain almost in isolation and lets the two triple bonds respond as independent oscillators. The insertion of Au therefore symmetrizes the vibrational coupling of the two C-C-C triple-bond units of the primitive unit cell, in addition to softening the average frequency.

Together, the structural, electronic, vibrational, and mechanical data collected so far converge on a coherent physical picture of the free-standing $\gamma$-graphdiyne molecular wires, addressing several questions left open by the on-surface synthesis of Cartoceti \textit{et al.}~\cite{Cartoceti2026}. The first concerns the microscopic identity of the structural and spectroscopic fingerprints reported on Au(100). The relaxed COW periodicity of $9.4464$~\AA{} per primitive cell reproduces the experimental scanning-tunneling-microscopy periodicity of approximately $0.95$~nm to within $-0.5\%$, anchoring the polymeric structural model to the experimental observable, and the relaxed OMW antisymmetric $\mathrm{C}{\equiv}\mathrm{C}$ frequency of $2081.9$~cm$^{-1}$ reproduces the experimental Raman peak near $2080$~cm$^{-1}$ to within $5$~cm$^{-1}$. The species-resolved mode assignment above adds a quantitative interpretation that the experiment alone cannot isolate, since the antisymmetric $\mathrm{C}{\equiv}\mathrm{C}$ mode carries no Au character in its eigenvector and the $-115$~cm$^{-1}$ shift from COW to OMW is therefore a purely electronic signature of $\mathrm{Au}(5d) \to \pi^{*}(\mathrm{C}{\equiv}\mathrm{C})$ back-donation and of the associated partial cumulenization of the chain.

The second question concerns the separation between intrinsic chain properties and substrate-induced renormalization. The phonon dispersion of the lattice-dynamical analysis above contributes to the only conclusion that the on-surface measurement cannot draw, since the experimental analysis of Cartoceti \textit{et al.}\ is restricted to zone-center modes of substrate-decoupled fragments and cannot separate the intrinsic stability of the isolated chain from the stabilizing effect of the Au(100) substrate. The full $\Gamma \to \mathrm{X}$ dispersion of both polymers in their free-standing form shows no imaginary modes within the finite-displacement floor of $0.4$~cm$^{-1}$, demonstrating that neither phase requires the substrate to achieve dynamic stability. The vibrational fingerprints characterized in the species-resolved mode assignment above, in particular the ECC mode at $1408$~cm$^{-1}$ specific to COW and the Au-projected modes below $400$~cm$^{-1}$ specific to OMW, are accordingly intrinsic and survive a moderate renormalization by the substrate, providing a vibrational basis for discrimination of the two phases beyond the $\mathrm{C}{\equiv}\mathrm{C}$ shift alone.

The third question concerns the electronic ground state of the OMW polymer. The four ground-state probes investigated in the equilibrium-geometry analysis above and the HSE06 refinements of the electronic-structure analysis above converge on a single answer once the band structure is resolved along a dense $\Gamma \to \mathrm{X}$ path rather than estimated from a moderate Monkhorst--Pack mesh. Closed-shell calculations, at PBE+MBD or HSE06, place the OMW polymer in the metallic regime with vanishing gap, and the same outcome is obtained with the HSE06 with a Gaussian-smeared 1-unit cell. The HSE06 calculation with collinear ferromagnetic initialization in a 1-unit primitive cell, which stabilizes a local spin-symmetry-broken solution, also turns out to be metallic when the dispersion is sampled with $101$ wave vectors, with the spin-$\alpha$ band crossing the Fermi level at $k_z \approx 0.23 \cdot 2\pi/c$ and the spin-$\beta$ band at $k_z \approx 0.27 \cdot 2\pi/c$, both with fractional occupation $f \approx 0.5$ at the crossing. The OMW polymer is therefore a one-dimensional metal under hybrid-functional treatment, with no Peierls dimerization instability accessible in the harmonic phonon spectrum (the lattice-dynamical analysis above) and with the structural and vibrational signatures of partial cumulenization reported in the structural and vibrational discussions above.

A complementary feature of the OMW electronic ground state, beyond the spin-resolved Fermi crossings of the periodic chain, concerns the finite vs. infinite crossover between OMW oligomers and the OMW polymer. The HSE06 electronic gap envelope $\{0.69, 0.35, 0.48, 0.37, 0\}$~eV for $n = 2, 3, 4, 6$ and the polymer is clearly non-monotonic on the finite side and closes to a true metallic state in the periodic limit, in contrast with the smooth Kuhn-type evolution characteristic of polyene and polyyne chains, and the standard fit $E_g(n) = a + b/(n + c)$ that summarizes that family is incompatible with the present data. The non-monotonicity on the finite side reflects the parity of the number of Au bridges along the oligomer, which is $n - 1$ and accordingly even for $n = 3$ and odd for $n = 2, 4, 6$. The odd-Au oligomers are open-shell $\pi$-radicals with a doublet ground state whose unpaired electron is distributed over the carbon backbone with a small admixture on Au, and their HSE06 gap is the SOMO-LUMO separation of the localized radical. The even-Au oligomer ($n = 3$) is closed-shell with no available radical to localize, and its HSE06 gap reflects the bare conjugation length of the $\pi$ system. The growth of the radical delocalization with $n$ then competes with the standard $1/n$ closing of the conjugation gap, producing the alternating envelope of finite oligomers and culminating, at the polymeric limit, in a Bloch state that delocalizes along the periodic chain and crosses the Fermi level, washing out the SOMO-LUMO gap of the finite limit into the spin-resolved Fermi crossings of the electronic-structure analysis above. The data accordingly trace a continuous chemical narrative between the molecular radical limit and the periodic metal that does not reduce to a monotonic Kuhn fit, and that situates the radical-to-metal transition of OMW as the central electronic feature distinguishing this phase from its polyynic COW counterpart.

Beyond the specific system, these observations carry a methodological message transferable to the broader class of $\pi$-conjugated wires functionalized with single-atom metal bridges: the PBE+MBD failure on the $n = 4$ and $n = 6$ multi-Au radicals discussed above is the expected manifestation of the GGA self-interaction error~\cite{Cohen2008,MoriSanchez2008} for this class, and the present finite vs. infinite OMW series quantifies on a chemically meaningful test set the extent to which HSE06 exact-exchange admixture remedies it.

\section{Conclusions}

We have presented a first-principles characterization of the free-standing $\gamma$-graphdiyne molecular wire in its organometallic (OMW) and covalent (COW) phases, combining PBE+MBD geometry optimizations, HSE06 hybrid functional electronic structure, Phonopy lattice dynamics, species-resolved mode decomposition, and a three-layer strain protocol applied to the relaxed polymeric reference. The COW polymer is a one-dimensional semiconductor with a direct electronic band gap of $E_g^\mathrm{HSE06} = 2.02$~eV at $\Gamma$ and a polyynic bond pattern that reproduces the experimental scanning-tunneling-microscopy periodicity of approximately $0.95$~nm. The OMW polymer is a partially cumulenized one-dimensional metal whose dense-path HSE06 dispersion shows the spin-$\alpha$ and spin-$\beta$ channels of the collinear ferromagnetic solution crossing the Fermi level at $k_z \approx 0.23 \cdot 2\pi/c$ and $k_z \approx 0.27 \cdot 2\pi/c$ respectively, with the relaxed antisymmetric $\mathrm{C}{\equiv}\mathrm{C}$ stretch at $2082$~cm$^{-1}$ reproducing the experimental Raman fingerprint of the on-surface phase. Both polymers are dynamically stable as free-standing objects, the Young's moduli are $995$~nN and $861$~nN respectively, with the $\mathrm{Au{-}C}$ bond identified as the softest mechanical link and absorbing more than half of the axial deformation in OMW, and species-resolved decomposition identifies the Au-projected modes below $400$~cm$^{-1}$ and the effective-conjugation-coordinate mode at $1408$~cm$^{-1}$ as vibrational fingerprints specific to OMW and COW respectively.

A methodological lesson emerges from the present analysis and is transferable to the broader class of $\pi$-conjugated wires functionalized with single-atom metal bridges, namely that PBE+MBD systematically over-delocalizes the unpaired density of multi-Au radicals and erases the parity-dependent doublet states of OMW oligomers at $n = 4$ and $n = 6$, which HSE06 recovers via local spin-symmetry breaking. The OMW polymer is, therefore, a one-dimensional metal in its free-standing form, with the spin-resolved Fermi crossings of the broken-symmetry HSE06 solution resolved by a dense $\Gamma \to \mathrm{X}$ band path with fractional-occupation analysis.


\section*{CRediT authorship contribution statement}
A.G.L.R.: Software, Validation, Formal Analysis, Data Curation, and Writing Original Draft. G.S.L.F.: Investigation, Software, Validation, and Writing Original Draft. B.I.: Investigation, Software, Validation, and Writing Original Draft. F.L.L.M.: Supervision, Resources, and Writing (Review and Editing). D.S.G.: Methodology, Supervision, Resources, Writing (Review and Editing), and Funding Acquisition. M.L.P.J.: Conceptualization, Methodology, Software, Validation, Formal Analysis, Investigation, Resources, Data Curation, Visualization, Writing (Review and Editing), Supervision, Project Administration, and Funding Acquisition.

\section*{Declaration of competing interest}
The authors declare that they have no known competing financial interests or personal relationships that could have appeared to influence the work reported in this paper.

\section*{Data availability}
The data that support the findings of this study are available from the corresponding author upon reasonable request.

\section*{Acknowledgements}
G.S.L.F.\ acknowledges the S\~ao Paulo Research Foundation (FAPESP) fellowship (process number 2024/03413-9). B.I.\ thanks CNPq and the S\~ao Paulo Research Foundation (FAPESP) (process numbers 194155/2025-0 and 2024/11016-0). D.S.G.\ acknowledges the Center for Computing in Engineering and Sciences at Unicamp for financial support through the FAPESP CEPID Grant (process number 2013/08293-7) and support from INEO/CNPq and FAPESP (grant 2025/27044-5). M.L.P.J.\ acknowledges financial support from FAPDF (grant 00193-00001807/2023-16), CNPq (grants 444921/2024-9 and 308222/2025-3), and CAPES (grant 88887.005164/2024-00).

\bibliographystyle{elsarticle-num}
\bibliography{references}

@article{Hirsch2010,
  author  = {Hirsch, Andreas},
  title   = {The era of carbon allotropes},
  journal = {Nature Materials},
  volume  = {9},
  pages   = {868--871},
  year    = {2010},
  doi     = {10.1038/nmat2885}
}

@article{Novoselov2004,
  author  = {Novoselov, K. S. and Geim, A. K. and Morozov, S. V. and Jiang, D. and Zhang, Y. and Dubonos, S. V. and Grigorieva, I. V. and Firsov, A. A.},
  title   = {Electric Field Effect in Atomically Thin Carbon Films},
  journal = {Science},
  volume  = {306},
  number  = {5696},
  pages   = {666--669},
  year    = {2004},
  doi     = {10.1126/science.1102896}
}

@article{Kroto1985,
  author  = {Kroto, H. W. and Heath, J. R. and O'Brien, S. C. and Curl, R. F. and Smalley, R. E.},
  title   = {{C$_{60}$}: Buckminsterfullerene},
  journal = {Nature},
  volume  = {318},
  pages   = {162--163},
  year    = {1985},
  doi     = {10.1038/318162a0}
}

@article{Iijima1991,
  author  = {Iijima, Sumio},
  title   = {Helical microtubules of graphitic carbon},
  journal = {Nature},
  volume  = {354},
  pages   = {56--58},
  year    = {1991},
  doi     = {10.1038/354056a0}
}

@article{Baughman1987,
  author  = {Baughman, R. H. and Eckhardt, H. and Kertesz, M.},
  title   = {Structure-property predictions for new planar forms of carbon: Layered phases containing {sp$^2$} and {sp} atoms},
  journal = {The Journal of Chemical Physics},
  volume  = {87},
  number  = {11},
  pages   = {6687--6699},
  year    = {1987},
  doi     = {10.1063/1.453405}
}

@article{Huang2018,
  author  = {Huang, Changshui and Li, Yongjun and Wang, Ning and Xue, Yurui and Zuo, Zicheng and Liu, Huibiao and Li, Yuliang},
  title   = {Progress in Research into {2D} Graphdiyne-Based Materials},
  journal = {Chemical Reviews},
  volume  = {118},
  number  = {16},
  pages   = {7744--7803},
  year    = {2018},
  doi     = {10.1021/acs.chemrev.8b00288}
}

@article{Xue2018,
  author  = {Xue, Yurui and Li, Yongjun and Zhang, Jing and Liu, Zhongfan and Zhao, Yuliang},
  title   = {{2D} Graphdiyne Materials: Challenges and Opportunities in Energy Field},
  journal = {Science China Chemistry},
  volume  = {61},
  pages   = {765--786},
  year    = {2018},
  doi     = {10.1007/s11426-018-9270-y}
}

@article{LiYL2014,
  author  = {Li, Yongjun and Xu, Liang and Liu, Huibiao and Li, Yuliang},
  title   = {Graphdiyne and graphyne: from theoretical predictions to practical construction},
  journal = {Chemical Society Reviews},
  volume  = {43},
  pages   = {2572--2586},
  year    = {2014},
  doi     = {10.1039/C3CS60388A}
}

@article{LiYL2010,
  author  = {Li, Guoxing and Li, Yuliang and Liu, Huibiao and Guo, Yanbing and Li, Yongjun and Zhu, Daoben},
  title   = {Architecture of graphdiyne nanoscale films},
  journal = {Chemical Communications},
  volume  = {46},
  pages   = {3256--3258},
  year    = {2010},
  doi     = {10.1039/B922733D}
}

@article{Casari2016,
  author  = {Casari, C. S. and Tommasini, M. and Tykwinski, R. R. and Milani, A.},
  title   = {Carbon-atom wires: {1D} systems with tunable properties},
  journal = {Nanoscale},
  volume  = {8},
  pages   = {4414--4435},
  year    = {2016},
  doi     = {10.1039/C5NR06175J}
}

@article{Tykwinski2010,
  author  = {Tykwinski, R. R. and Chalifoux, W. and Eisler, S. and Lucotti, A. and Tommasini, M. and Fazzi, D. and Del Zoppo, M. and Zerbi, G.},
  title   = {Toward carbyne: Synthesis and stability of really long polyynes},
  journal = {Pure and Applied Chemistry},
  volume  = {82},
  number  = {4},
  pages   = {891--904},
  year    = {2010},
  doi     = {10.1351/PAC-CON-09-09-04}
}

@article{Pan2011,
  author  = {Pan, L. D. and Zhang, L. Z. and Song, B. Q. and Du, S. X. and Gao, H.-J.},
  title   = {Graphyne- and graphdiyne-based nanoribbons: Density functional theory calculations of electronic structures},
  journal = {Applied Physics Letters},
  volume  = {98},
  number  = {17},
  pages   = {173102},
  year    = {2011},
  doi     = {10.1063/1.3583507}
}

@article{Bai2011,
  author  = {Bai, H. and Zhu, Y. and Qiao, W. and Huang, Y.},
  title   = {Structures, stabilities and electronic properties of graphdiyne nanoribbons},
  journal = {RSC Advances},
  volume  = {1},
  pages   = {768--775},
  year    = {2011},
  doi     = {10.1039/C1RA00481F}
}

@article{Bu2012,
  author  = {Bu, H. and Zhao, M. and Zhang, H. and Wang, X. and Xi, Y. and Wang, Z.},
  title   = {Isoelectronic doping of graphdiyne with boron and nitrogen: Stable configurations and band gap modification},
  journal = {The Journal of Physical Chemistry A},
  volume  = {116},
  number  = {15},
  pages   = {3934--3939},
  year    = {2012},
  doi     = {10.1021/jp300107d}
}

@article{Zhou2011,
  author  = {Zhou, J. and Lv, K. and Wang, Q. and Chen, X. S. and Sun, Q. and Jena, P.},
  title   = {Electronic structures and bonding of graphyne sheet and its {BN} analog},
  journal = {The Journal of Chemical Physics},
  volume  = {134},
  number  = {17},
  pages   = {174701},
  year    = {2011},
  doi     = {10.1063/1.3583476}
}

@article{Kang2012,
  author  = {Kang, J. and Wu, F. and Li, J.},
  title   = {Modulating the bandgaps of graphdiyne nanoribbons by transverse electric fields},
  journal = {Journal of Physics: Condensed Matter},
  volume  = {24},
  number  = {16},
  pages   = {165301},
  year    = {2012},
  doi     = {10.1088/0953-8984/24/16/165301}
}

@article{Zhu2016,
  author  = {Zhu, Y. and Bai, H. and Huang, Y.},
  title   = {Electronic Property Modulation of One-Dimensional Extended Graphdiyne Nanowires from a First-Principle Crystal Orbital View},
  journal = {ChemistryOpen},
  volume  = {5},
  number  = {1},
  pages   = {78--87},
  year    = {2016},
  doi     = {10.1002/open.201500154}
}

@article{Yang2008,
  author  = {Yang, S. and Kertesz, M.},
  title   = {Linear {C$_n$} clusters: Are they acetylenic or cumulenic?},
  journal = {The Journal of Physical Chemistry A},
  volume  = {112},
  number  = {1},
  pages   = {146--151},
  year    = {2008},
  doi     = {10.1021/jp076805b}
}

@article{Cinquanta2011,
  author  = {Cinquanta, E. and Ravagnan, L. and Castelli, I. E. and Cataldo, F. and Manini, N. and Onida, G. and Milani, P.},
  title   = {Vibrational characterization of dinaphthylpolyynes: A model system for the study of end-capped {sp} carbon chains},
  journal = {The Journal of Chemical Physics},
  volume  = {135},
  number  = {19},
  pages   = {194501},
  year    = {2011},
  doi     = {10.1063/1.3660355}
}

@article{Ravagnan2009,
  author  = {Ravagnan, L. and Manini, N. and Cinquanta, E. and Onida, G. and Sangalli, D. and Motta, C. and Devetta, M. and Bordoni, A. and Piseri, P. and Milani, P.},
  title   = {Effect of Axial Torsion on {sp} Carbon Atomic Wires},
  journal = {Physical Review Letters},
  volume  = {102},
  number  = {24},
  pages   = {245502},
  year    = {2009},
  doi     = {10.1103/PhysRevLett.102.245502}
}

@article{Cai2010,
  author  = {Cai, Jinming and Ruffieux, Pascal and Jaafar, Rached and Bieri, Marco and Braun, Thomas and Blankenburg, Stephan and Muoth, Matthias and Seitsonen, Ari P. and Saleh, Moussa and Feng, Xinliang and M\"ullen, Klaus and Fasel, Roman},
  title   = {Atomically precise bottom-up fabrication of graphene nanoribbons},
  journal = {Nature},
  volume  = {466},
  pages   = {470--473},
  year    = {2010},
  doi     = {10.1038/nature09211}
}

@article{Bjork2013,
  author  = {Bj{\"o}rk, Jonas and Hanke, Felix and Stafstr{\"o}m, Sven},
  title   = {Mechanisms of Halogen-Based Covalent Self-Assembly on Metal Surfaces},
  journal = {Journal of the American Chemical Society},
  volume  = {135},
  number  = {15},
  pages   = {5768--5775},
  year    = {2013},
  doi     = {10.1021/ja400304b}
}

@article{Bjork2016,
  author  = {Bj{\"o}rk, Jonas},
  title   = {Reaction mechanisms for on-surface synthesis of covalent nanostructures},
  journal = {Journal of Physics: Condensed Matter},
  volume  = {28},
  number  = {8},
  pages   = {083002},
  year    = {2016},
  doi     = {10.1088/0953-8984/28/8/083002}
}

@article{Klappenberger2018,
  author  = {Klappenberger, F. and Hellwig, R. and Du, P. and Paintner, T. and Uphoff, M. and Zhang, L. and Lin, T. and Moghanaki, B. A. and Paszkiewicz, M. and Vobornik, I. and Fujii, J. and Fuhr, O. and Zhang, Y.-Q. and Allegretti, F. and Ruben, M. and Barth, J. V.},
  title   = {Functionalized Graphdiyne Nanowires: On-Surface Synthesis and Assessment of Band Structure, Flexibility, and Information Storage Potential},
  journal = {Small},
  volume  = {14},
  number  = {14},
  pages   = {1704321},
  year    = {2018},
  doi     = {10.1002/smll.201704321}
}

@article{DeBoni2020a,
  author  = {De Boni, F. and Merlin, G. and Sedona, F. and Casalini, S. and Seyyed Fakhrabadi, M. M. and Sambi, M.},
  title   = {Templating Effect of Different Low-Miller-Index Gold Surfaces on the Bottom-Up Growth of Graphene Nanoribbons},
  journal = {ACS Applied Nano Materials},
  volume  = {3},
  number  = {11},
  pages   = {11497--11509},
  year    = {2020},
  doi     = {10.1021/acsanm.0c02596}
}

@article{Sedona2020,
  author  = {Sedona, F. and Seyyed Fakhrabadi, M. M. and Carlotto, S. and Mohebbi, E. and De Boni, F. and Casalini, S. and Casarin, M. and Sambi, M.},
  title   = {On-surface synthesis of extended linear graphyne molecular wires by protecting the alkynyl group},
  journal = {Physical Chemistry Chemical Physics},
  volume  = {22},
  number  = {21},
  pages   = {12180--12186},
  year    = {2020},
  doi     = {10.1039/D0CP01634A}
}

@article{DeBoni2024,
  author  = {De Boni, F. and Pilot, R. and Milani, A. and Ivanovskaya, V. V. and Abraham, R. J. and Casalini, S. and Pedron, D. and Casari, C. S. and Sambi, M. and Sedona, F.},
  title   = {Structure and vibrational properties of {1D} molecular wires: From graphene to graphdiyne},
  journal = {Nanoscale},
  volume  = {16},
  pages   = {11211--11222},
  year    = {2024},
  doi     = {10.1039/D4NR00943F}
}

@article{Rabia2020,
  author  = {Rabia, A. and Tumino, F. and Milani, A. and Russo, V. and Li Bassi, A. and Bassi, N. and Lucotti, A. and Achilli, S. and Fratesi, G. and Manini, N. and Onida, G. and Sun, Q. and Xu, W. and Casari, C. S.},
  title   = {Structural, Electronic, and Vibrational Properties of a Two-Dimensional Graphdiyne-Like Carbon Nanonetwork Synthesized on {Au(111)}: Implications for the Engineering of {sp-sp$^2$} Carbon Nanostructures},
  journal = {ACS Applied Nano Materials},
  volume  = {3},
  number  = {12},
  pages   = {12178--12187},
  year    = {2020},
  doi     = {10.1021/acsanm.0c02665}
}

@article{Achilli2021,
  author  = {Achilli, S. and Milani, A. and Fratesi, G. and Tumino, F. and Manini, N. and Onida, G. and Casari, C. S.},
  title   = {Graphdiynes interacting with metal surfaces: First-principles electronic and vibrational properties},
  journal = {2D Materials},
  volume  = {8},
  number  = {4},
  pages   = {044014},
  year    = {2021},
  doi     = {10.1088/2053-1583/ac26ad}
}

@article{Cartoceti2026,
  author  = {Cartoceti, A. and Achilli, S. and D'Agosta, P. and Tumino, F. and Garg, S. and Orbelli Biroli, A. and Onida, G. and Fratesi, G. and Russo, V. and Li Bassi, A. and Maier, S. and Casari, C. S.},
  title   = {Surface dependent organometallic to covalent transition in graphdiyne molecular wires},
  journal = {Nanoscale},
  volume  = {18},
  pages   = {336--350},
  year    = {2026},
  doi     = {10.1039/D5NR01968K}
}

@article{Perdew1996,
  author  = {Perdew, John P. and Burke, Kieron and Ernzerhof, Matthias},
  title   = {Generalized Gradient Approximation Made Simple},
  journal = {Phys. Rev. Lett.},
  volume  = {77},
  number  = {18},
  pages   = {3865--3868},
  year    = {1996},
  doi     = {10.1103/PhysRevLett.77.3865}
}

@article{Tkatchenko2012,
  author  = {Tkatchenko, Alexandre and DiStasio, Robert A., Jr. and Car, Roberto and Scheffler, Matthias},
  title   = {Accurate and Efficient Method for Many-Body van der {Waals} Interactions},
  journal = {Phys. Rev. Lett.},
  volume  = {108},
  number  = {23},
  pages   = {236402},
  year    = {2012},
  doi     = {10.1103/PhysRevLett.108.236402}
}

@article{Ambrosetti2014,
  author  = {Ambrosetti, Alberto and Reilly, Anthony M. and DiStasio, Robert A., Jr. and Tkatchenko, Alexandre},
  title   = {Long-range correlation energy calculated from coupled atomic response functions},
  journal = {J. Chem. Phys.},
  volume  = {140},
  number  = {18},
  pages   = {18A508},
  year    = {2014},
  doi     = {10.1063/1.4865104}
}

@article{Blum2009,
  author  = {Blum, Volker and Gehrke, Ralf and Hanke, Felix and Havu, Paula and Havu, Ville and Ren, Xinguo and Reuter, Karsten and Scheffler, Matthias},
  title   = {Ab initio molecular simulations with numeric atom-centered orbitals},
  journal = {Comput. Phys. Commun.},
  volume  = {180},
  number  = {11},
  pages   = {2175--2196},
  year    = {2009},
  doi     = {10.1016/j.cpc.2009.06.022}
}

@article{Krukau2006,
  author  = {Krukau, Aliaksandr V. and Vydrov, Oleg A. and Izmaylov, Artur F. and Scuseria, Gustavo E.},
  title   = {Influence of the exchange screening parameter on the performance of screened hybrid functionals},
  journal = {J. Chem. Phys.},
  volume  = {125},
  number  = {22},
  pages   = {224106},
  year    = {2006},
  doi     = {10.1063/1.2404663}
}

@article{Togo2015,
  author  = {Togo, Atsushi and Tanaka, Isao},
  title   = {First principles phonon calculations in materials science},
  journal = {Scripta Materialia},
  volume  = {108},
  pages   = {1--5},
  year    = {2015},
  doi     = {10.1016/j.scriptamat.2015.07.021}
}

@article{Cohen2008,
  author  = {Cohen, Aron J. and Mori-S{\'a}nchez, Paula and Yang, Weitao},
  title   = {Insights into Current Limitations of Density Functional Theory},
  journal = {Science},
  volume  = {321},
  number  = {5890},
  pages   = {792--794},
  year    = {2008},
  doi     = {10.1126/science.1158722}
}

@article{MoriSanchez2008,
  author  = {Mori-S{\'a}nchez, Paula and Cohen, Aron J. and Yang, Weitao},
  title   = {Localization and Delocalization Errors in Density Functional Theory and Implications for Band-Gap Prediction},
  journal = {Phys. Rev. Lett.},
  volume  = {100},
  number  = {14},
  pages   = {146401},
  year    = {2008},
  doi     = {10.1103/PhysRevLett.100.146401}
}

@article{Desyatkin2022,
  author  = {Desyatkin, Victor G. and Martin, William B. and Aliev, Ali E. and Chapman, Nathaniel E. and Fonseca, Alexandre F. and Galv{\~a}o, Douglas S. and Miller, Ericka Roy and Stone, Kevin H. and Wang, Zhong and Zakhidov, Dante and Limpoco, F. Ted and Almahdali, Sarah R. and Parker, Shane M. and Baughman, Ray H. and Rodionov, Valentin O.},
  title   = {Scalable Synthesis and Characterization of Multilayer $\gamma$-Graphyne, New Carbon Crystals with a Small Direct Band Gap},
  journal = {J. Am. Chem. Soc.},
  volume  = {144},
  number  = {39},
  pages   = {17999--18008},
  year    = {2022},
  doi     = {10.1021/jacs.2c06583}
}

@article{Aliev2025,
  author  = {Aliev, Ali E. and Guo, Yongzhe and Fonseca, Alexandre F. and Razal, Joselito M. and Wang, Zhong and Galv{\~a}o, Douglas S. and Bolding, Claire M. and Chapman-Wilson, Nathaniel E. and Desyatkin, Victor G. and Leisen, Johannes E. and Ribeiro Junior, Luiz A. and Kanegae, Guilherme B. and Lynch, Peter and Zhang, Jizhen and Judicpa, Mia A. and Parra, Aaron M. and Zhang, Mengmeng and Gao, Enlai and Hu, Lifang and Rodionov, Valentin O. and Baughman, Ray H.},
  title   = {A planar-sheet nongraphitic zero-bandgap $sp^2$ carbon phase made by the low-temperature reaction of $\gamma$-graphyne},
  journal = {Proc. Natl. Acad. Sci. U.S.A.},
  volume  = {122},
  number  = {5},
  pages   = {e2413194122},
  year    = {2025},
  doi     = {10.1073/pnas.2413194122}
}

@article{Kone2025,
  author  = {Kone, Ezra R. and Nasri, Sarah and Parker, Grace L. and Desyatkin, Victor G. and Martin, William B. and Trant, John F. and Rodionov, Valentin O.},
  title   = {Replication of mechanochemical syntheses of $\gamma$-graphyne from calcium carbide fails to produce the claimed product},
  journal = {Carbon},
  volume  = {232},
  pages   = {119808},
  year    = {2025},
  doi     = {10.1016/j.carbon.2024.119808}
}

@article{Martin2024,
  author  = {Martin, William B. and Warburton, Robert E. and Parker, Shane M. and Rodionov, Valentin O.},
  title   = {On the characterization of $\gamma$-graphyne},
  journal = {Nat. Synth.},
  volume  = {3},
  number  = {10},
  pages   = {1208--1211},
  year    = {2024},
  doi     = {10.1038/s44160-024-00642-1}
}

\end{document}